\documentclass[12pt]{article}

\usepackage{epsf}
\usepackage[dvips]{graphicx}

\begin{document}
\begin{center}
{\bfseries Formation of the intermediate baryon systems in
hadron-nuclear and nuclear-nuclear interactions}

\vskip 5mm

M. K. Suleymanov$^{1,2\dag}$, E. U. Khan$^{1}$, A Kravchakova$^{3}$
, Mahnaz Q. Haseeb$^{1}$,S. M. Saleem$^{1}$, Y. H.
Huseynaliyev$^{1,4}$, S Vokal$^{2,3}$, A.S. Vodopianov$^{2}$, O.B.
Abdinov$^{5}$

\vskip 5mm

{\small (1) {\it Department of Physics COMSATS Institute of
Information Technology, Islamabad, Pakistan }
\\
(2) {\it Veksler and Baldin Laboratory of High-Energies, JINR,
Dubna, Russia}
\\
(2) {\it  University of P. J. Shafarik, Koshice, Slovak Republic}
\\
(2) {\it Department of Semiconductor Physics, Sumgayit State
University, Azerbaijan}
\\
(2) {\it Physics Institute of AS, Baku, AzerbaijanVeksler and Baldin
Laboratory of High-Energies, JINR, Dubna, Russia}
\\
$\dag$ {\it E-mail: mais@jinr.ru}}
\end{center}

\vskip 5mm

\begin{center}
\begin{minipage}{150mm}
\centerline{\bf Abstract}The centrality experiments indicate regime
change and saturation in the behavior of some characteristics of the
secondary particles emitted in hadron-nuclear and nuclear-nuclear
interactions at high energies. The phenomenon has a critical
character.  The simple models do not explain the effect. We suppose
that the responsible mechanism to explain the phenomenon could be
the formation and decay of the intermediate baryon systems. Such
systems could be formed as a result of nucleon percolation in
compressed baryonic matter. Formation of big percolation cluster may
change the properties of the medium, e.g., it could lead to the
changing its transparency. This could be used to get a signal of the
intermediate baryonic system formation. We consider two signals to
identify the formation of the intermediate baryon systems:  the
critical changing of transparency of the strongly interacting matter
and the enhancement   of light nuclei production with increase in
centrality.
\end{minipage}
\end{center}

\vskip 10mm

\section{Centrality experiments}One of the important experimental methods to get  information on
the changes of states of nuclear matter by increasing its baryon
density is the study of characteristics of hadron-nuclear and
nuclear-nuclear interactions depending on the centrality of
collisions at high energies. On the other hand the centrality of
collisions can not be defined directly in the experiment. In
different experiments the values of  centrality  are defined as the
number of identified protons , projectiles'  and  targets'
fragments,  slow particles, all particles, as the energy flow of the
particles with emission  angles  $\theta = 0^0$   or  with
$\theta=90^0$ . Apparently, it is not simple to compare
quantitatively the results on centrality-dependences obtained in
literature while on the other hand the definition of centrality
could significantly influence the final results. May be this is a
reason, why we could not get a clear signal on new phases of
strongly interacting matter, though a lot of interesting information
has been given in those experiments. Let us consider some of the
experiments.

\subsection{Hadron - Nuclear Interactions} In Ref.~\cite{chemakin1} the results
from BNL experiment E910 on pion production and stopping in
proton-Be, Cu, and Au collisions as a function of centrality at a
beam momentum of 18 GeV/c are presented. The centrality of the
collisions is characterized using the measured number of "grey"
tracks, $N_{grey}$, and a derived quantity  , the number of
inelastic nucleon-nucleon scatterings suffered  by the projectile
during the collision. In Fig. 1 are plotted the values of average
multiplicity for $\pi^-$-mesons ($<\pi^-Multiplicity>$) as a
function of $N_{grey}$ and for the three different targets. We
observe that $<\pi^- Multiplicity>$ increases approximately
proportionally to $N_{grey}$  and for all three targets at small
values of $N_{grey}$ or $\nu$   and saturates with increasing
$N_{grey}$ and in the region of  higher values of $N_{grey}$  and
$\nu$ . Solid line in figure shows the expectations for the $<\pi^-
Multiplicity> (\nu )$ based on the wounded-nucleon (WN)
model~\cite{chemakin1} and with dashed lines, does a much better job
of describing p-Be yields than the WN model.

BNL E910 has measured $\Lambda$  production as a function of
collision centrality for 17.5 GeV/c p-Au collisions~\cite{ron1}.
Collision centrality is defined by $\nu$. The $\Lambda$ yield versus
$\nu$ is plotted in Fig. 2. The open symbols are the integrated
gamma function yields, and the errors shown represent $90\%$
confidence limits including systematic effects from the
extrapolations. The full symbols are the fiducial yields. These
various curves represent different functional scalings.
We see that the measured $\Lambda$ yield increases faster than the
participant scaling expectation for $\nu\le 3$ and then saturates.
The same result has been obtained by BNL E910 Collaboration for
$K^0_s$  and $K^+$ -mesons emitted in $p+Au$ reaction. Now let us
consider some example on nuclear-nuclear interactions.

\subsection{Nuclear-Nuclear Interaction} Fig. 3 presents the average
values of multiplicity $<n_s>$ for $s$ - particles produced in $Kr +
Em$ reactions at 0.95 GeV/nucl as a function of
centrality~\cite{abdinov2}. One can say that there are two regions
in the behavior of the values of $<n_s>$   as a function of $N_g$
for the Kr+Em reaction. In the region of: $N_g < 40$ the values of
$<n_s>$ increase linearly with $N_g$, here the cascade evaporation
model (CEM ~\cite{cem}) also gives the linear dependence but with
the slope less than the experimental one; $N_g  > 40$  CEM gives the
values for average $n_s$ greater than the experimentally observed
ones. The last saturates in this region and the effect could not be
described  the CEM. This has been already observed in emulsion
experiments~\cite{abduzhamilov1}.

\subsection{Heavy Ion Collisions} It is very important that the regime
change has been indicated in the behavior of heavy flavor particles
production in ultrarelativistic heavy ion collisions as a function
of centrality. The ratio of the $J/\Psi$  to Drell-Yan
cross-sections has been measured by NA38 and NA50 SPS CERN as a
function of the centrality of the reaction estimated, for each
event, from the measured neutral transverse energy
$E_t$~\cite{abreu1234}. Whereas peripheral events exhibit the normal
behaviour already measured for lighter projectiles or targets,
$J/\Psi$  shows a significant anomalous drop of about $20\%$ in the
$E_t$ range between 40 and 50 GeV. The detailed pattern of the
anomaly can be seen in Fig. 4 which shows the ratio of the $J/\Psi$
to the Drell-Yan cross-sections divided by the exponentially
decreasing function accounting for normal nuclear absorption.

           Other significant effect seen from this figure
           is a regime change in the   $E_t$ range between 40 and 50 GeV both
           for light and heavy ion collisions and saturation.

\section{Main Results and Discussion} At some values of centrality the
regime change and saturation appears  as a critical    phenomena for
hadron-nuclear, nuclear-nuclear interactions and heavy  ion collisions
in the range of energy from SIS up to RHIC almost for all particles
(from mesons, baryons,  strange particles up to charmonium).
        The simple models (such us WN and CEM) which are usually used
        to describe the high energy hadron-nuclear and nuclear-nuclear
        interactions could not explain the existence of the point of
        regime change  and saturation.
The results show that the dynamics of the phenomena should be same
for hadron-nuclear, nuclear-nuclear interactions and heavy ion
collisions independent of the energy and mass of the colliding
nuclei and the types of particles. The responsible mechanisms to
describe the above mentioned phenomena could be statistical and
percolation ones because of their critical character.
Ref.~\cite{claudia1} presented the complicated information about
using statistical and percolation models to explain the experimental
results coming from heavy ion physics. The regime change and
saturation was observed for hadron-nuclear and light nuclear-nuclear
interaction where it is very hard and practically impossible to
reach the necessary conditions to apply the statistical theory (the
statistical models have to give the more strong A-dependencies than
percolation mechanisms). Therefore, we believe that the responsible
mechanism to explain the phenomena could be the percolation cluster
formation ~\cite{satz1}-\cite{pajares1}. Big percolation cluster may
be formed in the hadron-nuclear, nuclear-nuclear and heavy ion
interactions independent on the colliding energy. But the structure
and the maximum values of the reaching density and temperature of
hadronic matter can be different for different interactions and may
depend on the colliding energy and masses in the framework of the
cluster. Ref.~\cite{satz2} discusses that deconfinement is expected
when the density of quarks and gluons becomes so high that it no
longer makes sense to partition them into color-neutral hadrons,
since these would strongly overlap. Instead we have clusters much
larger than hadrons, within which color is not confined;
deconfinement is thus related to cluster formation. This is the
central topic of percolation theory, and hence a connection between
percolation and deconfinement seems very
likely~\cite{satz3}-\cite{celik1}. So we can see that the
deconfinement could occur in the percolation cluster.
Ref.~\cite{satz2} explains the charmonium suppression as a result of
deconfinment in cluster too.

\section{Search for signal} Observation of the effects connected with formation and decay of the
percolation clusters in heavy ion collisions at ultrarelativistic
energies could be the first step for getting the information about
the onset stage of deconfinement. We consider two signals to
identify the formation of the intermediate baryon systems:

 - the critical change the transparency of strongly interacting matter;

- the enhancement of light nuclei production with the increasing
centrality.

\subsection{Critical change the transparency of strongly
interacting matter} The critical change of transparency can be
expected to influence the characteristics of secondary particles. As
collision energy increases, baryons retain more and more of the
longitudinal momentum of the initial colliding nuclei, characterized
by a flattening of the invariant particle yields over a symmetric
range of rapidities, about the center of mass - an indicator of the
onset of nuclear transparency. To confirm the deconfinement in
cluster it is necessary to study the centrality dependence in the
behavior of secondary particles yields and simultaneously, critical
increase in the transparency of the strongly interacting matter.
Appearance of the critical transparency could change the absorption
capability of the medium and we may observe a change in the heavy
flavor suppression depending on their kinematical characteristics.
It means that we have to observe the anomalous distribution of some
kinematical parameters because those particles which are from the
region with superconductive properties (from cluster) will be
suppressed less than the ones from noncluster area.  So, the study
of the centrality dependence of heavy flavor particle production
with fixed kinematical characteristics may give the information
about changing of absorption properties of medium. Comparison of
yields in different ion systems by using nuclear modification
factors such as $R_{CP}$ (involving Central and Peripheral
collisions) should provide information on the
hadronization~\cite{christelle1} . $R_{CP}$ highlighted the particle
type dependence at intermediate $p_T$ as it was suggested by
coalescence models~\cite{molnar1} leading to the idea that hadrons
result from the coalescence of quarks in the dense medium. At high
$p_T$, jet fragmentation becomes the dominant process to explain the
hadron formation. Hence, the quark constituents may be the relevant
degrees of freedom for the description of collision. Using the
relation $R_{CP}={n_1\over n_2}$(here e.g. $n_1$  and $n_2$ could be
heavy flavor particles yields with fixed values of $p_T$ and $\eta$
) as a function of centrality, the masses and energy it is possible
to get necessary information on the properties of the nuclear
matter. With such definition of the $R$ , appearance of transparency
could be identified and detected using the condition $R\simeq 1$ .
Using the statistical and percolation models~\cite{claudia1} and
experimental data on the behavior of the nuclear modification
factors one can get information on the appearance of the anomalous
nuclear transparency as a signal of formation of the percolation
cluster.

\subsection{Enhancement of a light nuclei production} There is a very positive chance that the effect of the light nuclei
emission~\cite{adler1}-\cite{bearden1} in heavy ion collisions are
to be one of the accompanying  effects of percolation cluster
formation and decay.  Light nuclei could be formed during the
formation of percolation cluster in pressure phase of nucleons
before deconfinement and in the phase of QGP expansion and cooling
as a result of nucleon coalescence. However, there is one more way
of light nuclei formation. It is well known that the light nuclei
can be formed mainly as a result of the disintegration of the
projectiles and the targets during the interaction. These processes
are called nuclear fragmentation and have been studied well.  The
yields of light nuclei (fragments) in this case have to increase
with centrality from peripheral collisions to semicentral ones then
the yields have to decrease as is shown in Fig.5 (taken form Ref.
~\cite{schuttauf1}).

In Fig.5 $Z_{bound} =\Sigma Z$ for fragments with $Z\ge 2$ emitted
in $Au+Au$ collisions at different energies. Obviously the yields of
the d, T, He and other light nuclei need to have almost similar
$Z_{bound}$ -dependencies. We find it is very interesting to study
another way of light nuclei formation, the formation as a result of
the recombination of created or stopped
nucleons~\cite{nagle1}(references therein). This recombination
process is called coalescence. The probability of coalescence of a
particular nuclear system depends on the properties of the hadronic
system formed as a result of the collision. So as an effect
accompanied by high density nuclear matter - the one under extreme
conditions, coalescence could provide us the information about the
states of high baryon density nuclear matter. Since the probability
of coalescence of a particular nuclear system depends on the
properties of the hadronic system formed as a result of the
collision, it may be expected that probability of nucleon
coalescence process could increase with the growing nuclear matter
density. Light nuclei are fairly large objects compared to simple
hadrons and their binding energies are small compared to freeze out
temperatures, which are of the order 100 MeV. These light clusters
are therefore not expected to survive through the high density
stages of the collision. The light nuclei observed in the experiment
are formed and emitted near freeze-out, and they mainly carry
information about this late stage of the collision. This is evident
from the simple nucleon coalescence
model~\cite{csernai1}-\cite{mekjian1}. Existing light nuclei
produced as a result of coalescence has to change the behavior of
the centrality dependences of light nuclei yields. The regime change
in the behavior of light nuclei yields as a centrality of collisions
is expected. So we believe that studying the yields of light nuclei
produced in the heavy ion collisions at relativistic and
ultrarelativistic energies as a function of collision centrality
could provide the information on formation of intermediate baryon
systems in hadron-nuclear and nuclear-nuclear interactions. In
experiment the light nuclei produced as a result of nucleon
coalescence mechanisms will be separated from other ones using the
following idea. The yields of the light nuclei produced as a result
of disintegration of the projectiles and the targets during the
interaction will behave as in Fig.5. Appearance of light nuclei
formed as a result of nucleon coalescence phenomenon can be a reason
of the regime change in the behavior of light nuclei yields.

It could be the first step of testing the idea and the second step
is to get some confirmation about the mechanism of the cluster
formation. To confirm that percolation mechanism could be a single
reason of the cluster formation it is necessary to show that the
regime change of the behavior of light nuclei yields as a function
of the centrality had a critical character and may be observed also
for the nuclei target (and projectile) with small atomic masses.

\section{Summary} Therefore the centrality experiments indicate the critical
appearance of the regime change and saturation in the behavior of
some characteristics of the secondary particles emitted in
hadron-nuclear and nuclear-nuclear interactions at high energies.
The underlining mechanism to explain the phenomena could be the
formation and decay of the intermediate baryon systems which may
form as a result of nucleon percolation in compression baryonic
matter. The critical changing of transparency of the strongly
interacting matter and the enhancement of the yield of light nuclei
production with centrality could be considered as two signals to
identify the formation of the intermediate baryon systems.

\begin{figure}[h]
 \centerline{
\includegraphics[width=120mm,height=80mm]{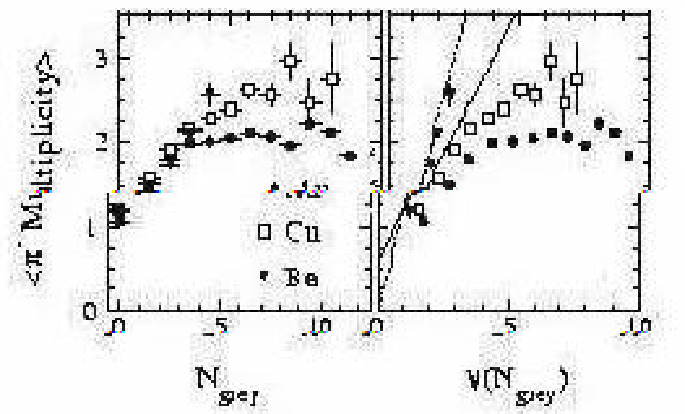}}
 \caption{The average multiplicity of the  $\pi^-$-mesons produced in
                  proton-Be, Cu, and Au collisions as a function of centrality  at
a beam momentum of 18 GeV/c. Solid line demonstrates the
                  results coming from the WN-model.}
\end{figure}

\begin{figure}[h]
 \centerline{
\includegraphics[width=80mm,height=60mm]{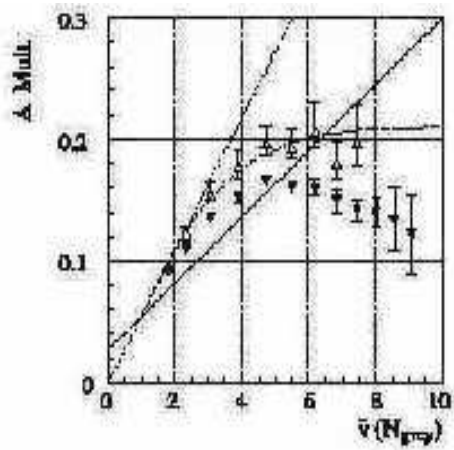}}
 \caption{The  $\Lambda$ yield versus $\nu$ (see text).}
\end{figure}

\begin{figure}[h]
 \centerline{
\includegraphics[width=60mm,height=60mm]{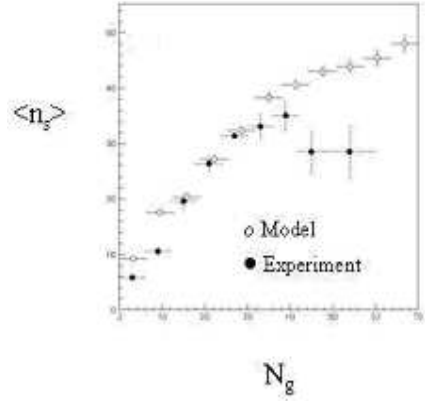}}
 \caption{The average values of multiplicity $<n_s>$ for $s$ -
            particles produced in $Kr + Em$ reactions at 0.95
            GeV/nucl as a function of centrality.}
\end{figure}
\begin{figure}[h]
 \centerline{
\includegraphics[width=80mm,height=70mm]{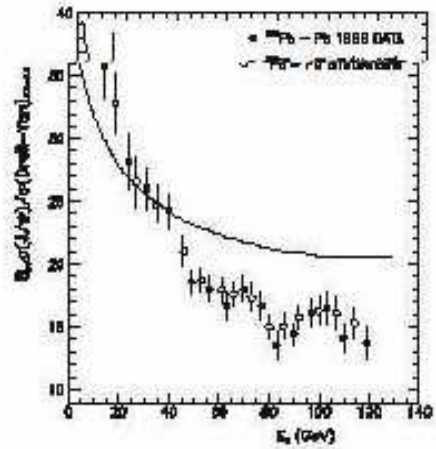}}
 \caption{The ratio of the $J/\Psi$  to the Drell-Yan cross-sections
                         divided by the exponentially decreasing  function
                         accounting for normal nuclear absorption.}
\end{figure}

\begin{figure}[h]
 \centerline{
\includegraphics[width=130mm,height=60mm]{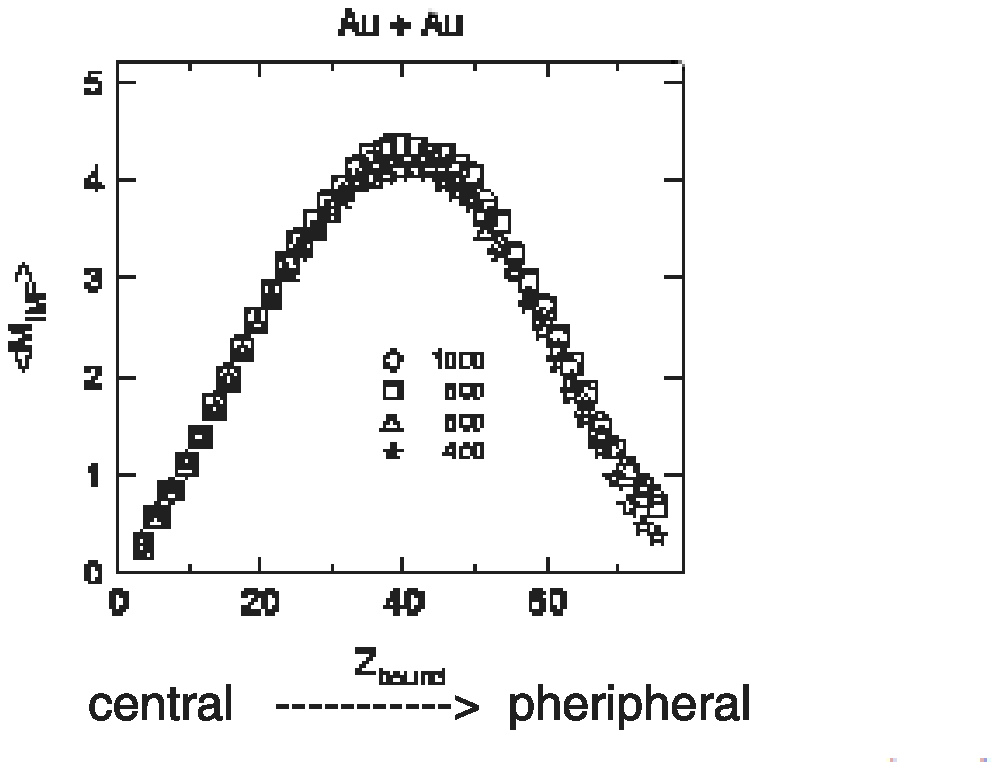}}
 \caption{}
\end{figure}

\end{document}